\documentclass[12pt]{article}
\usepackage{times}
\usepackage{geometry}
\usepackage{epsfig, epstopdf}
\usepackage[utf8]{inputenc}
\usepackage{enumitem,amssymb}
\usepackage{ragged2e}
\newlist{thematic}{itemize}{8}
\setlist[thematic]{label=$\square$}
\usepackage{pifont}
\usepackage{amsmath}
\usepackage{amssymb}

\newcommand{\simgt}{\lower 2pt \hbox{$\, \buildrel {\scriptstyle >}\over {\scriptstyle\sim}\,$}}
\newcommand{\simlt}{\lower 2pt \hbox{$\, \buildrel {\scriptstyle <}\over {\scriptstyle\sim}\,$}}
\newcommand{\msun}{\hbox{${M}_{\odot}$}}

\begin{document}

\def\sarc{$^{\prime\prime}\!\!.$}
\def\arcsec{$^{\prime\prime}$}
\def\beginrefer{\section*{References}%
\begin{quotation}\mbox{}\par}
\def\refer#1\par{{\setlength{\parindent}{-\leftmargin}\indent#1\par}}
\def\endrefer{\end{quotation}}

{\raggedright
\huge
Astro2020 Science White Paper \linebreak
\\
\noindent
Quasar Microlensing: Revolutionizing our Understanding of Quasar Structure and Dynamics \linebreak
\normalsize

\noindent \textbf{Thematic Areas:} \hspace*{60pt}  $\square$ Planetary Systems \hspace*{10pt} $\square$ Star and Planet Formation \hspace*{20pt}\linebreak
$\square$ Formation and Evolution of Compact Objects \hspace*{31pt} $\square$ Cosmology and Fundamental Physics \linebreak
  $\square$  Stars and Stellar Evolution \hspace*{1pt} $\square$ Resolved Stellar Populations and their Environments \hspace*{40pt} \linebreak
  $\text{\rlap{$\checkmark$}}\square$    Galaxy Evolution   \hspace*{45pt}  $\square$     Multi-Messenger Astronomy and Astrophysics \hspace*{65pt} \linebreak

\textbf{Principal Authors}\\
Timo Anguita (Universidad Andres Bello)\\
Leonidas Moustakas (Jet Propulsion Laboratory, California Institute of Technology)\\
Matthew O’Dowd (City University of New York and the American Museum of Natural History)\\
Rachel Webster (University of Melbourne)
\linebreak

\textbf{Co-authors:} 
George Chartas (College of Charleston), Matthew Cornachione (US Naval Academy), Xinyu Dai (University of Oklahoma), Carina Fian (Instituto de Astrofísica de Canarias and Departamento de
Astrofísica, Universidad de la Laguna), Damien Hutsemekers (University of Liege), Jorge Jimenez-Vicente (Univ. de Granada),
Kathleen Labrie (Gemini Observatory), Geraint Lewis (University of Sydney), Chelsea Macleod (Center for Astrophysics,
Harvard University), Evencio Mediavilla (Instituto de Astrofísica de Canarias),
Christopher W Morgan (US Naval Academy), Veronica Motta (Universidad de Valparaiso),
Anna Nierenberg (Jet Propulsion Laboratory), David Pooley (Trinity University),
Karina Rojas (Ecole Polytechnique Fédérale de Lausanne and LSSTC Data
Science Fellow), Dominique Sluse (STAR institute, University of Liège), Georgios
Vernardos (University of Groningen), Joachim Wambsganss (University of Heidelberg),
Suk Yee Yong (University of Melbourne)
 \linebreak

\textbf{Abstract:}}
Microlensing by stars within distant galaxies acting as strong
gravitational lenses of multiply-imaged quasars, provides a unique and direct
measurement of the internal structure of the lensed quasar on nano-arcsecond
scales. The measurement relies on the temporal variation of high-magnification
caustic crossings which vary on timescales of days to years. Multiwavelength
observations provide information from distinct emission regions in the quasar.
Through monitoring of these strong gravitational lenses, a full tomographic view can
emerge with Astronomical-Unit scale resolution. Work to date has demonstrated the
potential of this technique in about a dozen systems. In the 2020s there will be
orders of magnitude more systems to work with. Monitoring of lens systems for
caustic-crossing events to enable triggering of multi-platform, multi-wavelength
observations in the 2020s will fulfill the potential of quasar microlensing as a unique
and comprehensive probe of active black hole structure and dynamics.

\pagebreak

\begin{figure}
    \centering
  \includegraphics[width=0.85\textwidth]{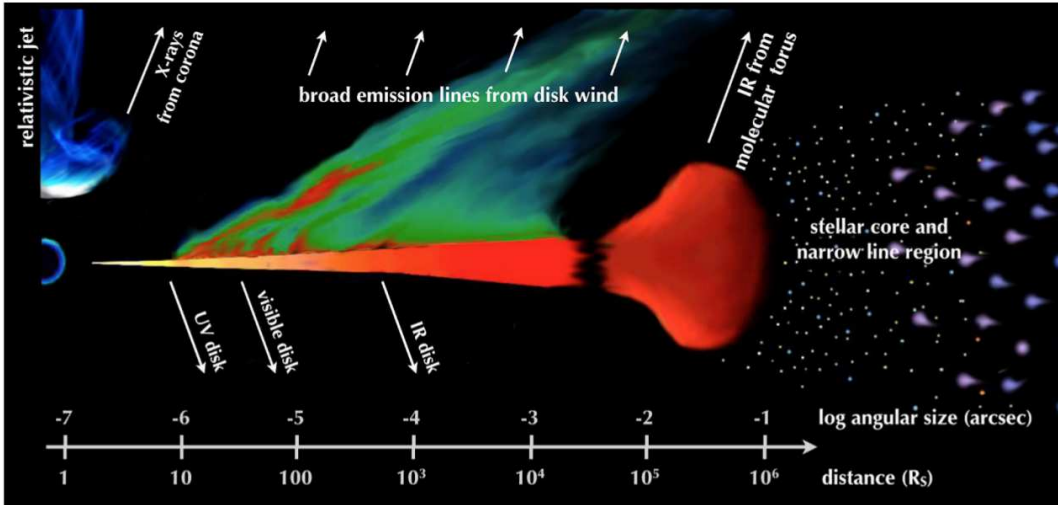}
     \caption{Quasar cross-section in log scale, with distance given in gravitational radii of central SMBH and log angular size for a typical luminous quasar at $z=1$ with black hole mass $10^8$\msun. (Adapted from images by Martin Elvis and Daniel Proga).
}
    \label{fig1}
\end{figure}

Quasars are powered by the accretion of matter onto a supermassive black hole
(SMBH; black holes with mass exceeding ~$10^6$\msun). Their central engines radiate via
a range of mechanisms that span several orders of magnitude in physical size, from
the inner thermal accretion disk and X-ray corona to the broad emission line region
(BELR) and relativistic jets, as shown in Figure~\ref{fig1}. At their cosmological distances,
we observe this complex structure on sub-microarcsecond scales, making direct
resolution an extraordinary challenge. Nevertheless, a detailed understanding of
quasar physics has profound scientific benefit: their emission processes can serve
as (often unique) laboratories for extreme physics; they have are one of our most
powerful probes of the early universe; and they are an important driver of galaxy
evolution and enrichment, cosmic star formation, and perhaps even reionization.

\begin{figure}
    \centering
  \includegraphics[width=0.9\textwidth]{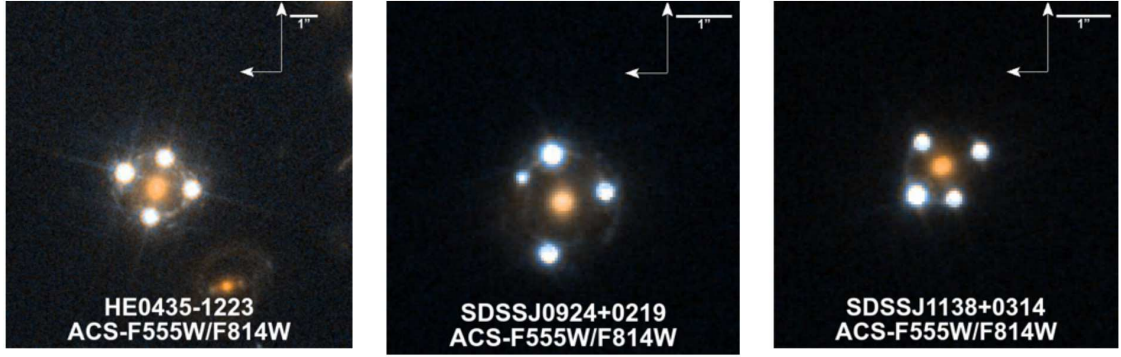}
     \caption{Hubble Space Telescope images of three representative four-image
strongly-lensed quasars, illustrating the types of objects that produce cosmological
microlensing. A 1\arcsec~ angular scale bar is shown in each panel. The intrinsic
variability of the lensed quasar will appear in each of the four images with a
time-delay that is set primarily by the gravitational profile of the lensing galaxy and
the relative distances. Each lensed image will feature additional magnification
changes through microlensing by the screen of stars within the lensing galaxy, along
each image’s line of sight. This is the signal discussed here.
}
    \label{fig2}
\end{figure}

\begin{figure}
    \centering
  \includegraphics[width=0.7\textwidth]{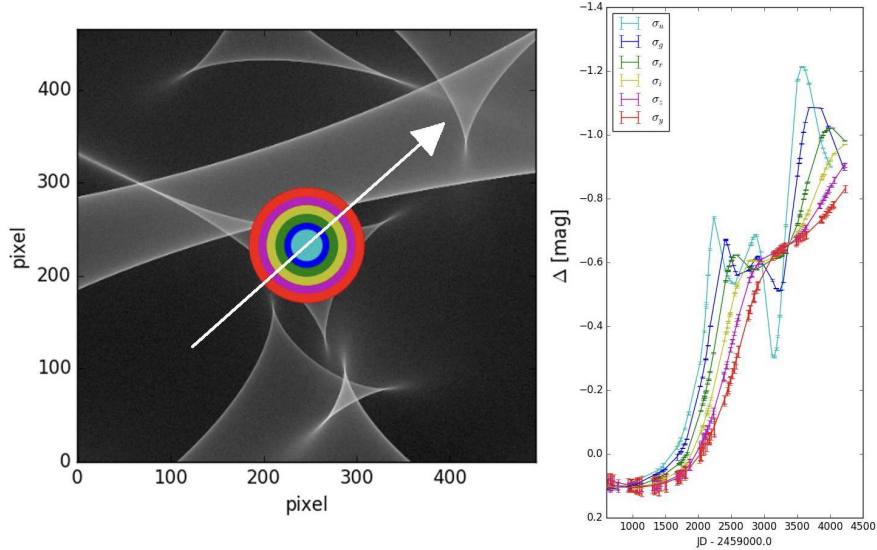}
     \caption{Adapted from Marshall+2017. Ten year long simulated light curve for
image A of RXJ1131-1231. The left panel shows the concentric Gaussian emission
regions observed by on the 6 LSST filters projected on top of the magnification
pattern. The right panel shows the extracted light curves interpolated at epochs
observed by LSST according to one (baseline) simulated observing strategy.
}
    \label{fig3}
\end{figure}

Microlensing of strongly-lensed quasars (Figures \ref{fig2} \& \ref{fig3}) provides us with a
technique capable of characterizing the internal structure of black hole environments
over orders of magnitude ranges of scale. In this paper we outline the motivation and
observational needs to fully exploit this technique. There are similarities in strategy to
what will be required for gravitational-wave multi-messenger follow-up observations,
and there may be an opportunity for coordinating the communities.

\smallskip
\smallskip
\noindent
{\bf Cosmological Microlensing:} Although the central engines of quasars are unresolveable by foreseeable telescopes, 
these structures may be tomographically
mapped when they are strongly-lensed into multiple images by a massive foreground
galaxy. The effect of individual stars within the lensing galaxy and along the line of
sight of each distinct lensed image of the quasar, results in a grainy but dramatic
substructure that varies on nano-arcsecond scales in the plane of the quasar and
that can lead to order unity or greater differential magnifications. This results in
size-dependent differential microlensing of the central engine. Furthermore, this is
not a static screen. Due to the relative transverse velocities of the quasar and
lensing galaxy from our point of view, as well as the general motion of the stars
within the lensing galaxy, these patterns constantly shift, resulting in secular
magnification changes in each lensed quasar image (Fig.~\ref{fig3}).

Using multi-wavelength observations of multiple caustic-crossing events, it is
possible in principle to reconstruct the two-dimensional structure of the emitting
regions corresponding to each wavelength. Due to the magnifications involved
(Fig.~\ref{fig3}), the angular sizes resolved correspond to physical scales of a few
Astronomical Units (Fig.~\ref{fig1}). With observations at corresponding wavelengths, it
becomes possible to measure the inner radius of the accretion disk, as well as its
overall structure and environment. These are related directly to the black hole mass
and spin. By identifying and tracking multiple such events in a large sample of
lensed quasars, it will be possible to address a series of fundamental questions
about the nature of accreting supermassive black holes, including:
\smallskip
\smallskip

$\bullet$ The mass and spin distribution of SMBHs;

$\bullet$ The relationship between SMBH mass and growth rate;

$\bullet$ Quasar accretion efficiency through the accretion disk temperature gradient;

$\bullet$ Whether re accretion disks are uniform, or exhibit structural complexity;

$\bullet$ Whether General Relativity is an accurate description of SMBHs.

\smallskip
\smallskip
In addition to the quasar physics, there is speculation that features in such systems
may be calibratable to known physical sizes or luminosities, in particular from the
dust-sublimation distance, which may become a new standard rod or standard
candle. There is already tantalizing work in this direction (e.g. Lusso+Risaliti 2017).

\smallskip
\smallskip
\noindent
{\bf Insights to Date:} Since the discovery of the first gravitational lens in 1979
(Walsh+1979) of the two-image lensed quasar Q0957+561, there have been
numerous concerted efforts to monitor such systems. Microlensing by the
lens-galaxy stars is routinely detected (Irwin+1989, Anguita+2008, Morgan+2010),
and indeed is a systematic that must be accounted for when the primary goal of
monitoring is to measure time delays between images towards Hubble constant
measurements (Eigenbrod+2005, Suyu+2017). To date, the following main
breakthroughs in quasar physics measurements that have been accomplished:

\smallskip
\smallskip
$\bullet$ Constraints on some accretion disk sizes and temperature gradients revealing
possible tension with the standard Shakura-Sunyaev profile (e.g.
Blackburne+2011, Bate+2018);

$\bullet$ Constraints on the geometry and kinematics of a broad-emission line region
(e.g. O’Dowd+2015, Sluse+2012);

$\bullet$ Measurements of X-ray corona geometry (e.g. Morgan+2008, Dai+2010).

\smallskip
\smallskip
Our community recognizes that the cadence, photometric precision, spectroscopic
information and wavelength range of the observations possible to date are not up to
the challenge of disentangling the full complexity of microlensing signals, but that
these are all areas that can be tackled this coming decade, with great promise.
The gold-standard microlensing experiment is an high-cadence, multi-wavelength
observation of a quasar under a caustic-crossing event--the transit of an
high-magnification caustic due to an individual star across the background source.
The nano-arcsecond gradient of such a curve scans the central engine structure with
resolution on the scale of the SMBH event horizon. Observation of several such
events would produce a rich dataset for mapping quasar structure. In practice,
quasars are variable objects, and the intrinsic variability itself depends on
wavelength. By virtue of the multiple imaging in these systems, however, the same
overall monitoring campaign required to trigger caustic-crossing event observations
will provide the difference in arrival time for light between each lensed image, so that
the intrinsic-variation behavior can be separated from the caustic-crossing
microlensing magnification changes. Photometric and spectroscopic monitoring can
also map emission line time delays which can be used for reverberation mapping -- a
complementary measurement. Interpretation of reverberation mapping time delays
relies on a suitable physical model for the quasar (Peterson 1993).
The greatest challenge to the micro-lensing experiment is the timescales involved.
The median Einstein crossing time (where ~1 caustic-crossing event is expected)
within a single lensed quasar image is approximately once every 20 years
(Mosquera \& Kochanek 2011). Even the most dramatic overachiever known today,
“Huchra’s Lens” Q2237+0305 with an unusually high caustic-crossing frequency (up
to 6 per decade considering all four lensed images; Wyithe+2000) has eluded
comprehensive follow-up due to the challenges of traditional monitoring. For this
reason, efforts to date have relied on statistical inference based on single- or
few-epoch observations during high-magnification events, and photometric
light-curves of low-magnification fluctuations.

\smallskip
\smallskip
\noindent
{\bf What the 2020s Holds:} Today, approximately a few dozen bright four-image
lensed-quasar systems are known. By the mid-2020s, due to both ground- and
space-based wide-field surveys including Dark Energy Survey, DESI, LSST and
Euclid , at least hundreds , perhaps 1000s of bright and robustly modellable systems
will be found (e.g. Oguri \& Marshall 2010). LSST will monitor the entire southern
observable sky with a cadence of about 80 times per annum (Ivezic+2018), giving us
a decade of “free” monitoring of these lensed quasars (Fig.~\ref{fig2}). The rate of
detected caustic-crossing events is expected to be in the hundreds per year, many
for bright sources.

The quasar microlensing community is faced with a powerful set of challenges if we
are to make optimal use of this upcoming flood of data. These challenges include:
1) Selection of a target sample of lensed quasars, to create a statistically significant
set of sources that will represent a meaningful span of properties such as black hole
mass, luminosity, orientation, and internal kinematic behavior. With a large enough
sample, it would be desirable to represent multiple cosmic look-back times, as well.
2) Development of a practical set of target-of-opportunity (TOO) targets drawn from
the full sample above, and follow-up programs, both imaging and spectroscopic,
spanning the EM spectrum. We must also ensure that the necessary facilities,
instruments, and observing configurations are in place. Such resources will likely be
increasingly available to respond to gravitational wave multi-messenger events. The
microlensing program can supplement and complement such infrastructure.
3) Computational facilities and analytic tools must be in place to deal with these data.
Of particular importance is the development of real-time light-curve analysis
techniques that give us absolute confidence in our TOO triggers for caustic crossing
events, and which also to derive maximum information from the full 10-year
light-curves.
4) Human resources must be in place to ensure that these opportunities are fully
realized.

\smallskip
\smallskip
\noindent
{\bf Conclusions:} Gravitational microlensing of multiply-imaged quasars has proved its
potential to resolve the nano-arcsecond structure of quasar central engines. The
ultimate promise of the technique is full tomographic mapping of their structure.
However, to achieve this goal, a statistically large number of quasars must be
monitored through high-magnification caustic-crossing events. The next generation
of wide-field survey is expected to bring this promise to fruition as they detect up to
hundreds of such events per year. However, a number of challenges must be met in
order to take full advantage of this opportunity. We must: understand which quasars
and events are desirable candidates for follow-up; prepare proposals and
instrumentation for multi-platform follow-up spanning the EM spectrum; develop
computational techniques and facilities to analyze a huge and complex data set in
real time; recruit and train the personnel needed for this challenging endeavor. If
these challenges can be met, then by the end of the 2020s we expect a detailed
geometric and kinematic model of quasar central engines spanning the full range of
emission properties. With this depth of understanding, quasars will become powerful
laboratories of extreme physics and even more powerful cosmological probes;
moreover their important role in cosmic evolution will be greatly clarified.

\pagebreak
\noindent
{\bf \Large References}\\
\\
\noindent
Anguita, T. et al. (2008), A\&A, 481, 615 \\
Bate, N.K. et al. (2018), MNRAS, 479, 4796 \\
Blackburne, J., A. et al. (2011), ApJ, 729, 34 \\
Dai, X. et al. (2010), ApJ, 709, 278 \\
Eigenbrod, A. et al. (2008), A\&A, 436, 25 \\
Irwin, M.J. et al. (1989), ApJ, 98, 1989 \\
Ivezic, Z., et al. (2018), 2008, arXiv0805.2366I \\
Lusso, E. \& Risaliti, G. (2017), A\&A, 602, L79 \\
Marshall et al. (2017), arXiv:1708.04058 and

https://github.com/LSSTScienceCollaborations/ObservingStrategy \\
Morgan, C. et al. (2008), ApJ, 689, 755 \\
Morgan, C. et al. (2010), ApJ, 712, 1129 \\
Mosquera, A. \& Kochanek, C (2011), ApJ, 738, 96 \\
O’Dowd, M.J., et al. (2015), ApJ, 813, 62 \\
Oguri, M. \& Marshall, P. (2010), 405, 2579 \\
Peterson, B.M. (1993), PASP, 105, 247 \\
Sluse, D., et al. (2012), A\&A, 544, 62 \\
Suyu, S. et al. (2017), MNRAS, 468, 2590 \\
Walsh, D., Carswell, R. F. \& Weymann, R. J. (1979), Nature, 279, 381 \\
Wyithe, J.S,B., Webster, R.L., \& Turner, E.L. (2000), MNRAS, 315, 337 \\

\end{document}